\RequirePackage{fix-cm}
\documentclass[twocolumn]{svjour3}          
\smartqed  
\usepackage{graphicx}

\newcommand{\echo}[0]{{EChO}}
\newcommand{\echosim}[0]{{ECHOSim}}
\newcommand{\micron}[0]{{$\mu$m}}

\journalname{Exp Astron}

\def\3he{$^3{\rm He}$}
\def\4he{$^4{\rm He}$}

\begin{document}

\title{\echosim: The Exoplanet Characterisation Observatory software simulator}


\author{E.~Pascale$^{1}$, I.~P.~Waldmann$^{2}$, C.~J.~MacTavish$^{3}$, A.~Papageorgiou~$^{1}$,
A.~Amaral-Rogers$^{1}$, R.~Varley$^{2}$, \\ 
V.~Coud\'e~de~Foresto$^{4}$, M.~J.~Griffin$^{1}$, 
M.~Ollivier$^{5}$, S.~Sarkar$^{1}$, L.~Spencer$^{6}$, B.~M.~Swinyard$^{2,7}$, M.~Tessenyi$^{2}$, G.~Tinetti$^{2}$  }


\institute{
$^{1}$ School of Physics and Astronomy, Cardiff University, Cardiff, UK \\
$^{2}$ Department of Physics and Astronomy, University College London,  London, UK\\
$^{3}$ Kavli Institute for Cosmology, University of Cambridge, Cambridge, UK \\ 
$^{4}$ Laboratoire d’Etudes Spatiales et d’Instrumentation en Astrophysique, Meudon, France\\
$^{5}$ Institut d’Astrophysique Spatiale, Université Paris XI, Orsay, France \\
$^{6}$ Department of Physics and Astronomy, University of Lethbridge,  Lethbridge, Alberta, Canada \\
$^{7}$ Space Science and Technology Department, Rutherford Appleton Laboratory, Didcot, UK 
}
\date{Received: date / Accepted: date}

\maketitle

\begin{abstract}
\echosim\ is the end-to-end time-domain simulator of the Exoplanet Characterisation Observatory (\echo) space mission. \echosim\ has been developed to assess the capability \echo\ has to detect and characterize the atmospheres of transiting exoplanets, and through this revolutionize the knowledge we have of the Milky Way and of our place in the Galaxy.
Here we discuss the details of the \echosim\ implementation and describe the models used to represent the instrument and to simulate the detection. 
Software simulators have assumed a central role in the design of new instrumentation and in assessing the level of systematics affecting the measurements of existing experiments.  
Thanks to its high modularity, \echosim\ can simulate basic aspects of several existing and proposed  spectrometers for exoplanet transits, including instruments on the Hubble Space Telescope and Spitzer, or ground-based and balloon borne experiments. A discussion of different uses of \echosim\  is given, including examples of simulations performed to assess the \echo\ mission.

\end{abstract}

\section{Introduction} 

The study of planets orbiting stars other than our Sun is one of the most fascinating and rapidly growing fields in the physical sciences. Rapidly growing also is the list of confirmed exoplanets, which currently exceeds one thousand, and is soon expected to reach tens of thousands of new alien worlds as  new instrumentation is being deployed on the ground and in space. Pioneering work in the last decade has allowed us to go beyond simple detection and to attempt the characterization of gaseous atmospheres. The technique used is transit spectroscopy where the signal of a transiting exoplanet's atmosphere superimposes a tiny modulation in time over the dazzling signal of the parent star during a transit or during an eclipse (\cite{Tinetti2007} \cite{Grillmair2008} \cite{Charbonneau2008}\cite{Swain2008b} \cite{Swain2008a} \cite{Pont2008} \cite{Swain2009} \cite{Thatte2010} \cite{Tinetti2010} \cite{Stevenson2010} \cite{Beaulieu2011} \cite{Mooij2011} \cite{Knutson2011} \cite{Sing2011} \cite{Bean2011} \cite{Brogi2012} \cite{Crouzet2012} \cite{Mooij2012} \cite{Deming2013} and \cite{tinetti2013}). Signals are small, at the level of $1\times 10^{-4}$ of the star, and detections therefore require an exquisite control of observational and instrumental systematics both at instrument level and during data analysis. The risk is in coupling the star signal into the signal of the exoplanet atmosphere. When observing from the 
ground, the Earth's atmospheric emission can also couple to the much smaller planet's signal if systematics are not sufficiently under control. This is not the first time we face such a problem, which can be effectively addressed only through dedicated instrumentation. The Exoplanet Characterisation Observatory~\cite{tinetti2012}, \echo, is a proposed space mission designed for photometric stability over a spectral band spanning from the visible to the mid-IR part of the electromagnetic spectrum and \echosim\ is the end-to-end software simulator developed to aid the instrument definition and to validate the mission concept. 

Static radiometric instrument models can assess the sensitivity of an experiment in delivering its scientific goals, but are often inadequate to model challenging systematic effects which can jeopardise the detection. This is particularly true when systematics manifest in the time-domain with a non-trivial temporal behaviour. For instance, the coupling between the stability of the telescope pointing and the focal plane detectors in the presence of non-perfect flat-fielding, or when detector pixel responses depart from a spatially flat optical transfer function.  The \echosim\ simulator has been developed to study the impact of this and other time-domain systematics, but it can also be used, as any radiometric model, to assess the overall sensitivity of the instrument. With a parametric definition of the mission concept, \echosim\ is used to validate the \echo\ mission concept and its ability to deliver the mission science requirements. Because of the high modularity, and execution efficiency, \echosim\ can 
also 
be used for a number of different applications such as the design of novel ground-based and balloon-borne exoplanet spectroscopic experiments, or to study the reliability of an existing detection by specifying a suitable instrument model in a parametric form. 

In this work, we review the algorithms implemented in \echosim\ and the approximations made to make the execution of these simulations efficient so that run times of the order of a few tens of seconds are achieved on normal laptop-size computers. The code is fully implemented in Python, with standard numerical libraries for portability, and simulations can be run on Windows, Linux and MacOS based machines. Each simulation begins with the generation of the frequency and time-dependent astronomical signal expected from the extrasolar system, which is then propagated to a cascade of processes that are involved in the instrument detection. 
In the final part of the paper we also briefly discuss possible uses of the simulator.

\section{\echo\ Instrument Design}
\begin{figure}[t]
  \centering
  \begin{minipage}[t][4.5cm][t]{0.4\textwidth}
    \includegraphics[width=\textwidth]{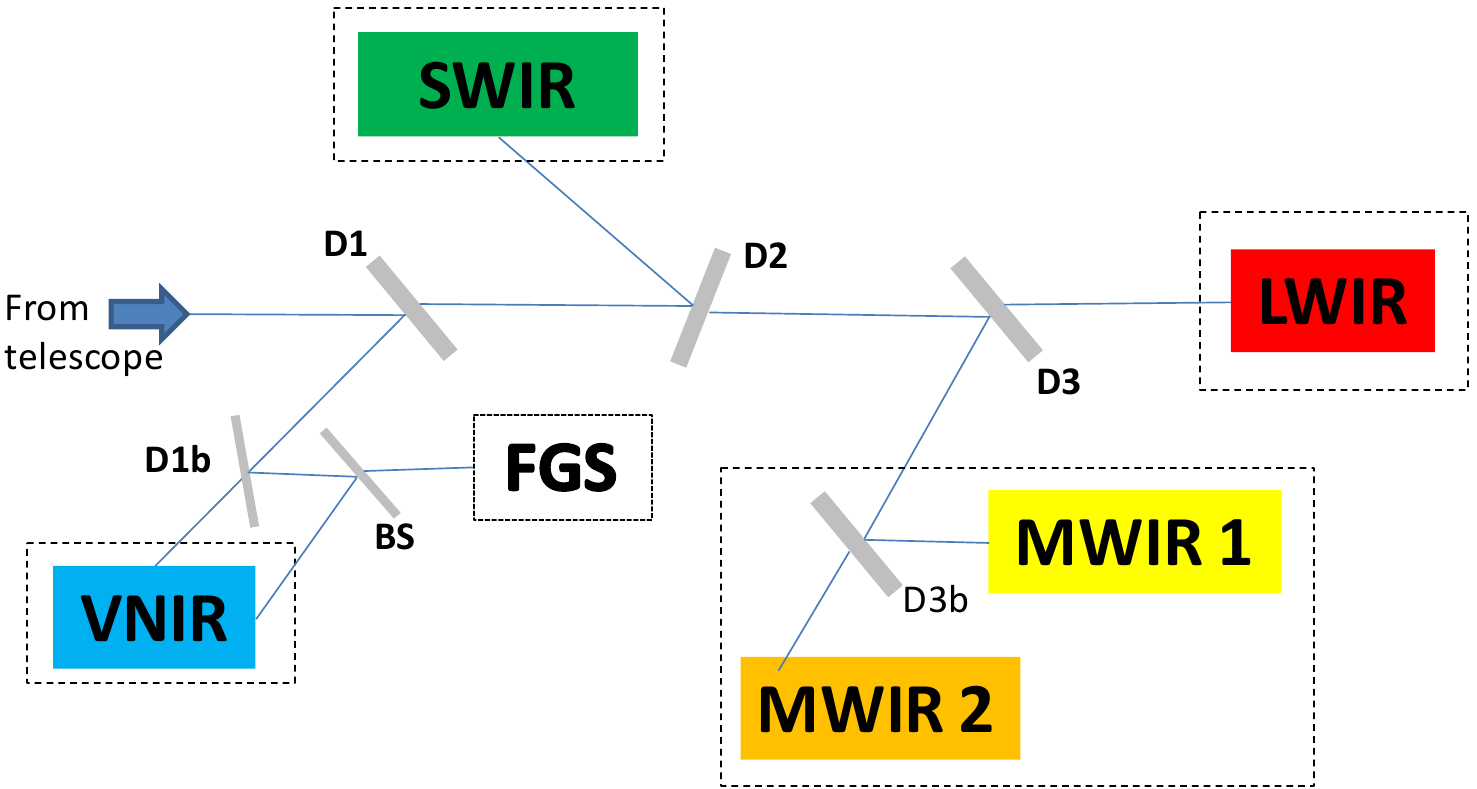}  
  \end{minipage}
  \begin{minipage}[b]{0.4\textwidth}
    \includegraphics[width=\textwidth]{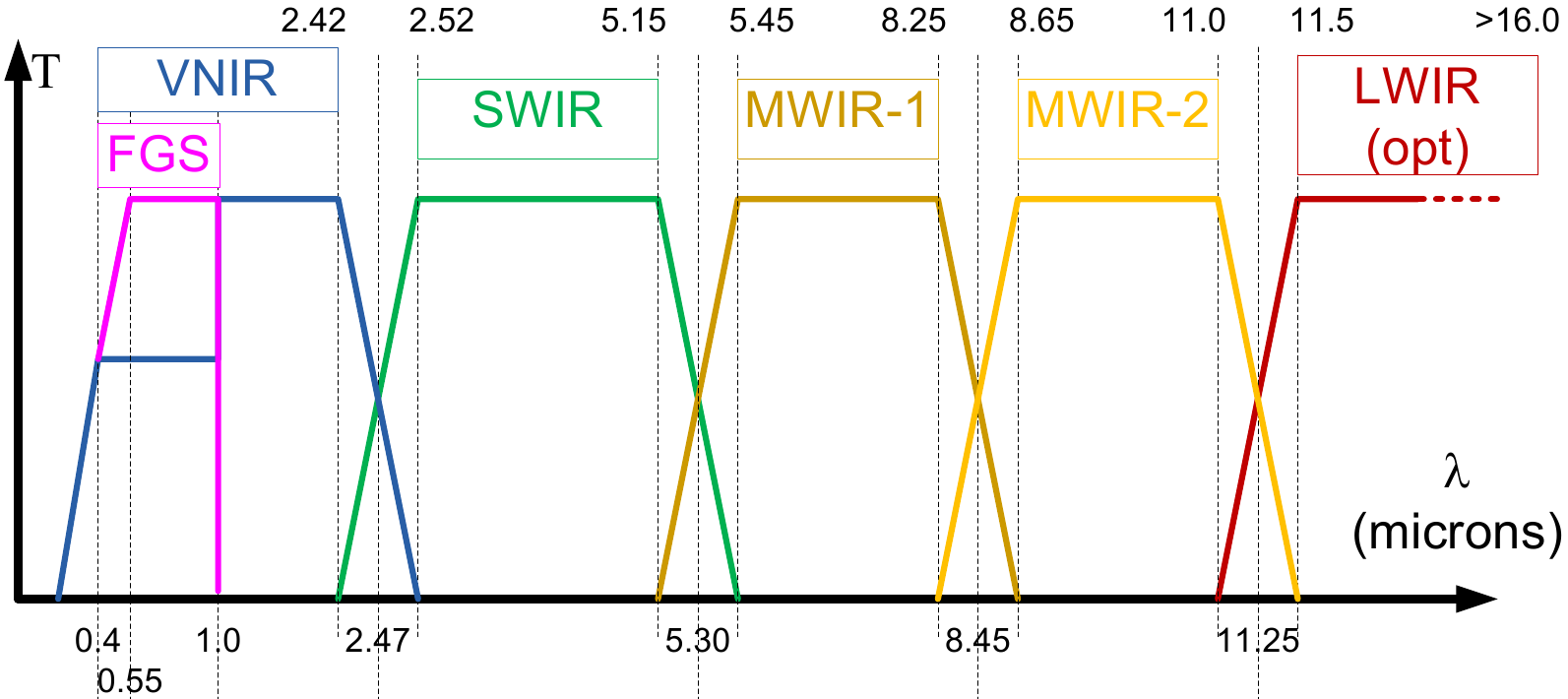}
   
  \end{minipage}
\caption{Top panel: baseline concept for the \echo\ payload channel separation. Bottom panel: \echo\ payload instrument channel division.}
\label{fig:1}       
\end{figure}

The \echo\ instrument is discussed briefly in this section. A more comprehensive review of this mission concept can be found in the \echo\ assessment study report: the ``yellow book''.\footnote{http://sci.esa.int/echo/53446-echo-yellow-book/.}
The telescope has a 1.2\,m aperture which is passively cooled to below 50\,K. The radiation collected by the primary aperture feeds five spectroscopic channels which cover continuously the required spectral band from 0.55 to 11\,\micron,  with the goal to extend the short and long wavelength ends to 0.4 and 16\,\micron, respectively. 
Because of the  large spectral coverage required, the \echo\ band is divided into five partially overlapping spectral channels, as is schematically shown in Figure \ref{fig:1}. 
The five channels are: the Visible and Near Infrared (VNIR) channel; the Short Wavelength Infrared (SWIR) channel; the Medium Wavelength Infrared (MWIR-1 and MWIR-2) channels; the Long Wavelength Infrared (LWIR) channel. An optical Fine Guidance System (FGS) is also part of the science payload. This is a star tracker  used in the attitude control system, and it is not implemented in \echosim.
The spectral band covered by each channel is schematically represented in Figure~\ref{fig:1}. 

Each spectral element is sampled by 2 detector pixels and the spectral resolving power is designed such that the requirements in the Mission Requirements Document\footnote{http://sci.esa.int/echo/51293-echo-mission-requirements-document/.} are satisfied. These are: resolving power greater than 300 at wavelengths under 5\micron, and resolving power greater than 30 at wavelengths above 5\micron. \echo\ is designed for photometric stability to be better than $10^{-4}$ over a period which can be 10\,hours long.

\section{The \echosim\ simulator}
The purpose of the \echosim\ simulator is to provide a software tool to assess all aspects of the \echo\ mission concept baseline, and alternative solutions, by using realistic, time-domain simulations of the astronomical scene and instrument. The instrument definition within the simulator is highly configurable, and it is possible to implement instruments other than \echo, including existing ground-based and space instruments.

\begin{figure}[t]
  \centering
    \includegraphics[width=0.4\textwidth]{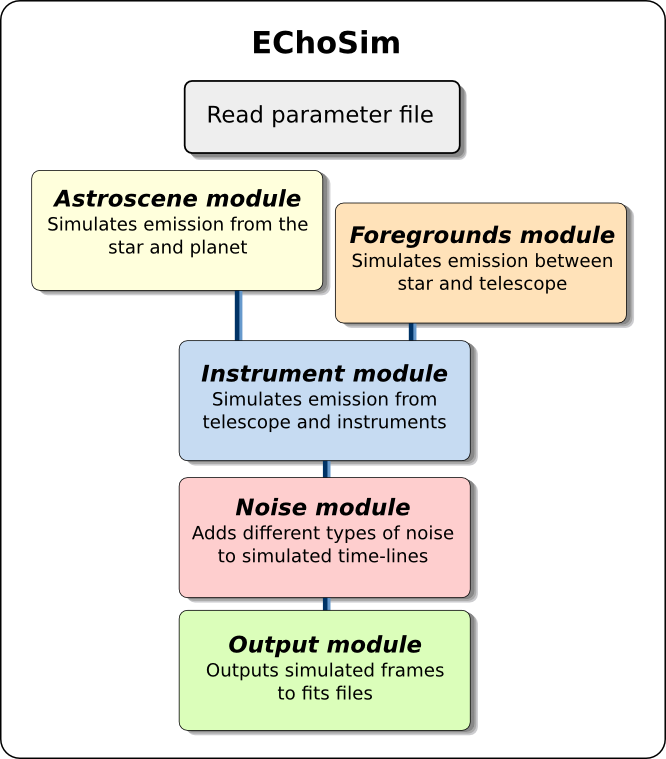} 

\caption{Overview of the \echosim\ architecture. The main software programme calls a number of modules describing the sky
and the payload instrument parameters.}
\label{fig:2}       
\end{figure}

The simulator design is optimized for computational efficiency without compromising the fidelity of the simulated detection. This allows each simulation to be run in seconds, opening up the possibility of Monte Carlo analyses of instrumental and astrophysical effects. For this reason, the simulator implementation is  highly modular (see Figure~\ref{fig:2}). A parametric description of the instrument and of the astronomical scene constitute the \echosim\ inputs. The Astroscene module simulates the extrasolar planetary system and computes wavelength dependent light curves. Zodiacal light computed by the Foreground module is superimposed on the exoplanet signal. For those cases where \echosim\ is used to assess ground-based or balloon-borne experiments, the Foreground module can use emission and transmission models of the Earth's atmosphere. The instrumental detection is simulated by the Instrument module which outputs noise-free detector timelines. The Noise module estimates sources of astrophysical and 
instrumental 
noise and superimposes a realization of the noise onto the outputs of the Instrument module. The final detector 
timelines (signal + noise) are written to disk (Output module) for subsequent analyses.

For computational efficiency, simulations are split into a slow and a fast temporal domain. The light curves are sampled on an irregular temporal grid with a cadence chosen to be the minimum required to Nyquist sample the time-varying astronomical signal. All quasi-static processes involving the detection and light dispersion on the focal planes are computed in this temporal domain. Faster processes include telescope pointing jitter, noise, detector acquisition, etc. These are computed at the end of the simulation in the Noise module, and co-added to the Instrument module's signals (re-sampled on a faster temporal grid, adequate to represent the sampling done by the detector pixels in the focal plane arrays).

\subsection{Astroscene module}
The observed flux from the combined star and planet is computed in a three step process involving
\begin{itemize}
 \item The stellar emission, calculated taking into account the star diameter, effective temperature and distance. The user has the option to approximate the flux using a black body spectrum. Alternatively,  Spectral Energy Distributions (SED) tabulated in a library precomputed using PHOENIX\footnote{http://www.hs.uni-hamburg.de/EN/For/ThA/phoenix/index.html.} stellar models can be used.
 
 \item The planetary contribution to the flux received by the telescope,  estimated for a primary transit or a secondary eclipse, as a wavelength-dependent transit depth, which accounts for the exoplanet's atmospheric emission, transmission, as well as for reflected stellar light. 
 
 \item The planetary orbit, simulated using the analytical description of \cite{mandel2002}. Light curves computed in this way are normalized taking into account the wavelength-dependent stellar SED and transit depth.
\end{itemize}

For primary transits, limb darkening effects are accounted for if the wavelength is smaller than 5\micron, but no limb darkening is assumed at longer wavelengths. For this purpose, quadratic limb darkening coefficients are taken from \cite{claret2000} and linearly interpolated over the the spectral band.
The transmission spectrum can be provided by the user, precomputed using a radiative transfer model. When this is not available \echosim\  estimates a wavelength-independent primary transit signal where the  depth of the light curve is given by 
\begin{equation}
 p = \frac{R_p}{R_s^2} \left(R_p + 2 \times 5H \right),
\end{equation}
where $R_p$ and $R_s$ are, respectively, the planet and star radii, and $H$ is the atmospheric scale height given by
\begin{equation}
 H = \frac{k_b T_p}{\mu g},
\end{equation}
where $k_b$ is the Boltzmann constant, $T_p$ is the planet's temperature, $g$ its gravity acceleration and $\mu$ the mean molecular weight of the atmosphere (see e.g. \cite{tessenyi2012}).

For secondary eclipses, the light curve's depth is estimated from both the planetary day-side emission and the reflected starlight.
The fraction of starlight reflected by the exoplanet is proportional to its geometric albedo and projected planetary surface area as a function of orbital phase.

The emission spectrum can be provided by the user as a wavelength-dependent planet-star contrast ratio, or be estimated assuming a  black body emission computed for the planet's temperature. In either case, the magnitude of the signal for the secondary eclipse is, in units of the stellar SED,
\begin{equation}
 p(\lambda, t) = \Phi(t) \frac{R_p^2}{R_s^2} 
  \frac{F_p(\lambda)}{F_s(\lambda)} ,
\end{equation}
where $F_p$ and $F_s$ are the planet and star SED, respectively, and $\Phi(t)$ is the exoplanet view factor (i.e. the fraction of the planetary day-side visible at a given orbital phase).

The exoplanet's temperature plays an important role in the simulations. When not supplied by the user, \echosim\ estimates the temperature using a simple radiative balance argument between the radiating energy received from the  star (accounting for albedo), and the  emission, which is assumed to be a black body. Inclusion of emissivity and whether the planet is tidally locked is also possible.

The transit depth estimated for the primary transit and secondary eclipse cases are then used to estimate the signal at the telescope. This is done by calculating the planet orbital solution and light curve~\cite{mandel2002},  normalized to physical units using the calculated transit depth and stellar SED. 

\subsection{Foregrounds module}
For space instrumentation operating at \echo's wavelengths, one major source of background emission is due to Zodiacal light. This is dominated at short wavelengths ($\lambda < 3.5$\micron) by scattered sunlight, and at longer wavelengths by the thermal emission from the same dust.
The zodiacal emission is implemented as a modified version of the JWST-MIRI Zodiacal model, pa\-ra\-me\-tri\-zed as
\begin{equation}
I_{\rm zodi}(\lambda) = B_\lambda (5500 K) 3.5 \times 10^{-14} + B_\lambda(270 K) 3.58 \times 10^{-8},  
\label{eq:zodi}
\end{equation}
where $B_\lambda (T)$ is the black body function. 
In order to represent the variation in Zodiacal light seen at different ecliptic latitudes, three regions of emission are defined. These correspond to minimum ($0.9 \times I_{Zodi}$), average ($2.5 \times I_{Zodi}$), and maximum ($8 \times I_{Zodi}$) emission. This is an adequate description for the exoplanet systems studied with the simulator. Exoplanets on a line of sight with higher Zodiacal column density are not considered, as extinction (in the visible) or foreground emission (in the IR) becomes prohibitively high for the detection.

The multiplicative factor of the Zodiacal model can also be user-defined to represent Zodiacal emission towards a particular target's line of sight. This is estimated from the best fit of 
equation \ref{eq:zodi} to a realistic Zodiacal model~\cite{kelsall1998}, obtained at the ecliptic 
latitude and longitude of the target, and at the expected time of observation. 

Because \echosim\ can simulate instruments other than \echo, Earth atmospheric emission and transmission models can be used for simulations of ground-based and balloon-borne experiments. Atmospheric spectra can be pre-computed by the user with Modtran\footnote{http://modtran5.com/.} or similar software and used in \echosim\ simulations. 

\subsection{Instrument module}

The detection of light is simulated in the instrument module. Light is collected by the telescope and then split by a series of dichroic filters to feed the 5 spectroscopic channels on \echo. Dispersion, diffraction and emission from optical surfaces and from instrument enclosures are all effects taken into account, as well as the detection by the focal plane detector pixels with their non-ideal response to light.

The telescope comprises a user-defined number of reflecting surfaces. For  \echo\ these are the primary, secondary and tertiary mirrors and a beam-folding mirror. Each of these surfaces have user-defined, wavelength-dependent reflectivities, used to estimate the overall telescope efficiency, $\eta_{\rm tel} = \eta_0 \prod_i r_i(\lambda)$, where the $r_i(\lambda)$ is the reflectivity of the $i$-th reflective surface and $\eta_0$ is an overall efficiency, estimated using optical cad software. 

\subsubsection*{Telescope}
The output of the simulated telescope comprises the signals from the point source (star and planet), the foregrounds and  the emission from the optical surfaces of the telescope, and are, respectively,
\begin{eqnarray}
   &&Q_{P,T}(\lambda, t) = \eta_{\rm tel} A_{\rm eff}  F_{\rm ps}(\lambda, t)  \\
   &&Q_{D, T} (\lambda)=  \eta_{\rm tel} A_{\rm eff} I_{\rm zodi} (\lambda) \\
   &&Q_{O,T} (\lambda) = \epsilon_N(\lambda) B_\lambda(T_N) + \nonumber \\
   && \hspace{2cm}+\sum_{i=1}^{N-1} \epsilon_i (\lambda) B_\lambda(T_i) \prod_{j = i+1}^N r_j(\lambda) ,
\end{eqnarray}
where $T_i$ is the temperature of each of the $N$ optical surfaces with user defined emissivities $\epsilon_i(\lambda)$. The effective area of the telescope is $A_{\rm eff}$, and  $F_{\rm ps}(\lambda, t)$ is the time-dependent flux from the star and the planet computed by the Astroscene module. These three signals are maintained in three different data structures because they behave differently when dispersed by the channel spectrometers. 

\subsubsection*{Dichroic filters}
The telescope output is split into channels by a set of dichroic filters. Each filter has user-defined transmission and reflection spectra. The emission of each filter is calculated from its (user-defined) wavelength-dependent emissivity  properties, in a way similar to the emission of the reflective elements of the telescope. The input to each spectroscopic channel are $Q_{P,C}(\lambda, t)$,  $Q_{D,C}(\lambda)$, and $Q_{O,C}(\lambda)$, which contain the effects on the light passing through the filters for the point source, diffuse radiation and optics emission, respectively.

\subsubsection*{Dispersive Optics and Diffraction}
Regardless of the technology used to disperse light in each channel, \echosim\ assumes that 
each focal plane has a spectral and a spatial axis. Light is dispersed along the spectral direction, the $x$-axis in a $xy$ reference located at the focal plane. The spatial direction of the spectrometer is along the $y$-axis.    
The spectral dispersion on the focal plane is given by a linear dispersion law between wavelength and position along the $x$-axis:
\begin{equation}
  LD = \frac{\Delta x}{\Delta\lambda} = {2\Delta_{pix}}\frac {R(\lambda_0)}{\lambda_0},
\end{equation}
where $\Delta_{pix}$ is the linear dimension of a detector pixel in the focal plane and $R(\lambda_0) = \lambda_0/\Delta\lambda_0$ is the spectral resolving power estimated at the central wavelength of each channel.  This is a very good approximation for the \echo\ baseline design as for many grating spectrometers with a spectral resolving power proportional to the wavelength.
There is a factor of 2 in the above equation to reflect a design where each spectral element, $\Delta\lambda$, is sampled by two detector pixels, required for Nyquist-sampling of spectral features.

The dispersed signals are sampled by the detector pixels assuming a diffraction pattern, or Point Spread Function (PSF). For the \echo\ instrument, this is approximated as a top-hat function for the  fibre-fed VNIR channel, and by a Gaussian function for the longer wavelength channels.
Assuming a diffraction limited instrument, the Gaussian PSF is
\begin{eqnarray}
  p(x, y, \lambda) =  \frac{1}{\sqrt{2\pi}\,\sigma_x} &&e^{-\left[ x - x_0(\lambda) \right]^2 / 2\sigma_x^2} \times \nonumber\\
		    &&\times \frac{1}{\sqrt{2\pi}\,\sigma_y} e^{-\left( y - y_0 \right)^2 / 2\sigma_x^2} ,
  \label{eq:psf}
\end{eqnarray}
where the coordinate $x_0$ is a function of the wavelength through the linear dispersion law
\begin{equation}
 x_0(\lambda) = LD \times (\lambda - \lambda_0).
\end{equation}
The size of the PSF for a diffraction limited instrument is
\begin{equation}
 \sigma_x = \frac{1}{\pi} \sqrt{2/K_x}\,F_\#\,\lambda{\rm ,}\;\;\;\sigma_y = \frac{1}{\pi} \sqrt{2/K_y}\,F_\#\,\lambda,
\end{equation}
where $F_\#$ is the telescope's $f$-number. The two constants $K_x$ and $K_y$ are used to model optical aberrations. One important aspect for computational efficiency is that the PSF is the product of two functions in the independent variables $x$ and $y$: $p(x,y,\lambda) = p(x,\lambda) p(y,\lambda)$. In addition to the diffraction limited and fibre-fed instrument  cases, \echosim\ can implement arbitrary PSFs provided from optical models in the form of two-dimensional illumination patterns. 

The PSF sampled by each detector is the convolution between the PSF and the intra-pixel response ($F(x)$ and $F(y)$), i.e. the real-valued optical transfer function of the detector pixel. This is given by
\begin{eqnarray}
 F(x) &=& \arctan\left\{\tanh\left[\frac{1}{2 l_d}\left(x + \frac{\Delta_{pix}}{2} \right) \right] \right\} + \nonumber\\ 
	&&\hspace{0cm}-\arctan\left\{\tanh\left[\frac{1}{2 l_d}\left(x - \frac{\Delta_{pix}}{2} \right) \right] \right\},
\end{eqnarray}
where the diffusion length $l_d$ is set to 1.7\micron\ (see \cite{barron2007} -- $F(y)$ has a similar expression in the $y$-coordinate). Therefore the PSF sampled by each detector pixel is the effective pixel response:
\begin{eqnarray*}
 &&p_s(x, y, \lambda) = p_s(x, \lambda) p_s(y, \lambda)\\
 && \hspace{1cm} p_s(x,\lambda) = p(x, \lambda) \ast F(x) \\ 
 && \hspace{1cm} p_s(y,\lambda) = p(y, \lambda) \ast F(y),
\end{eqnarray*}
where the correlation operator is indicated by the symbol ``$\ast$''.

\subsubsection*{Point Source}
The signal detected by each detector pixel in the focal plane array is the convolution between the incoming signal, $Q_{P,C}(\lambda, t)$, and the sampled PSF:
\begin{eqnarray*}
 &&Q_{P}(i,j,t) =  \\
  &&\hspace{0.5cm}\int {\rm QE}(\lambda)\; Q_{P, C}(\lambda, t) \;p_s\left[x(\lambda)-x_i, y_j, \lambda_i\right]\; d\lambda .
\end{eqnarray*}
The detector quantum efficiency is $QE(\lambda)$. The indices $i$ and $j$ identify the detector pixel located at physical coordinates $x_i$ and $y_j$ in the focal plane array. 
A relation between the pixel's physical coordinates, the indices and the wavelength exists:
\begin{eqnarray}
 && (x_i, y_j) = (i, j)\Delta_{pix} \\
 && \lambda = \frac{x}{LD} + \lambda_0 ,
\end{eqnarray}
where $i = -N_x/2 \dots (N_x/2 -1)$, $j = -N_y/2 \dots (N_y/2 -1)$, and $N_x$ and $N_y$ are the number of detector pixels in the spectral and spatial direction, respectively. The wavelength sampled at the centre of the $ij$-pixel is $\lambda_i = {i\Delta_{pix}}/{LD} + \lambda_0$.

For computational efficiency, \echosim\ takes advantage of the fact that the size of the sampled PSF changes less rapidly compared to a displacement in the coordinates. Therefore the spatial component of the sampled PSF can be taken out of the integral:
\begin{eqnarray*}
 Q_{P}(i,j,t) &&\simeq p_s(y_j, \lambda_i)\times \\
  && \times\int {\rm QE}(\lambda)\; Q_{P, C}(\lambda, t) \;p_s\left[x(\lambda)-x_i, \lambda_i\right]\; d\lambda .
 \label{eq:pointsource}
\end{eqnarray*}
With this approximation, the convolution between the spectrum and the PSF can be computed in one dimension, and the effects of the instrument's pointing stability (discussed later) can be studied independently on the spatial and spectral axes, when required.

\subsubsection*{Diffuse Radiation}
Diffuse radiation from the Zodiacal light (and the Earth's atmospheric emission when simulating sub-orbital experiments) contributes to the loading on a detector pixel. This is proportional to the pixel solid angle, 
\begin{equation}
  \Omega_p =  \left(\frac{\Delta_{pix}}{f_{\rm eff}} \right),
\end{equation}
where $f_{\rm eff}$ is the effective focal length. An input slit is used to limit the level of the background and a slit image is formed at the focal plane. If $L$ is the slit's linear dimension 
measured in the focal plane, then a given detector pixel receives diffuse radiation over the wavelength range
\begin{equation}
 \left( \lambda_i - \frac{\Delta_{pix}}{2}\frac{L}{LD}, \lambda_i + \frac{\Delta_{pix}}{2}\frac{L}{LD} \right) .
\end{equation}
Therefore the signal sampled by the $ij$-pixel is
\begin{equation}
 Q_{D}(i,j) = \Omega_p \int_{\lambda_i - \frac{\Delta_pix}{2}\frac{L}{LD}}^{\lambda_i + \frac{\Delta_pix}{2}\frac{L}{LD}}
		{\rm QE}(\lambda)\;Q_{D,C}(\lambda)\;d\lambda .
 \end{equation}

\subsubsection*{Instrument Emission}
The instrument emission sampled by a detector pixel depends on its entedue: $G = \frac{\pi}{4}\Delta_{pix}^2/f_\#$, where $f_\#$ is the working $f$-number. The signal sampled by the $ij$-pixel has a similar  expression to the  case of diffuse radiation discussed above:
\begin{equation}
 Q_{O}(i,j) = G \int_{\lambda_i - \frac{\Delta_pix}{2}\frac{L}{LD}}^{\lambda_i + \frac{\Delta_pix}{2}\frac{L}{LD}}
		{\rm QE}(\lambda)\;Q_{O,C}(\lambda)\;d\lambda .
 \end{equation}

\subsection{Noise module}
The Noise module simulates the main sources of instrumental and astrophysical noise. In a real instrument, noise sources act  at every stage of the detection, but \echosim\ implements  addition at the very end of each simulation. This is required for computational efficiency as the random processes of noise require a larger (temporal) bandwidth, which is very different from the bandwidth of the astronomical signal.  

The outputs from the instrument module simulations are signal-only timelines which are sampled with a cadence chosen to be the minimum required to Nyquist-sample the time-varying astronomical signal, i.e. the modulation in the light curve. The first task of the noise module is to re-sample the timelines to the user-defined detector sampling rate. The following noise components are then added to the signal-only timelines: 
\begin{itemize}
 \item Photon noise;
 \item Detector dark current and dark current noise;
 \item Detector readout noise;
 \item Detector inter-pixel gain variation;
 \item Telescope Pointing jitter.
\end{itemize}
These are all  independent random processes with the exception of the telescope pointing jitter which is correlated among all detector pixels in every focal plane. Although many of these processes have a Poisson distribution, they are all treated as Gaussian processes because they all have sufficiently large expected values. For instance, the dark current signal, $Q_{DC}(i,j,t)$ is a Gaussian process with mean dark current, $I_{DC}$,   standard deviation $\sigma_{DC} = \sqrt{I_{DC}}$, and $I_{DC} = I_0 e^{-\alpha/T_D}$ where $I_0$ and $\alpha$ are detector-specific parameters, and $T_D$ is the temperature of the focal plane array. 
The detector readout noise assumes a simple follow-up-the-ramp readout strategy and it depends on the number of non-destructive reads, $N_{\rm NDR}$
\begin{equation}
 \sigma_{\rm ro}^2 = 12 \frac{N_{\rm NDR}-1}{N_{\rm NDR}(N_{\rm NDR}+1)} \sigma_{\rm r}^2 ,
\end{equation}
where $\sigma_{\rm r}$ is the 1--$\sigma$  readout noise on a single non-destructive read. 

Detector gains are expected to be measured both before launch and during operations, and are used by the data reduction pipeline to flat-field the focal plane array when reconstructing the exoplanet's spectrum. \echosim\ implements pixel-to-pixel detector gain variation (or inter-pixel response) to simulate uncertainties in the flat-field operation. A detector-specific gain, $G_{ij}$, is randomly assigned to each pixel with a user-defined RMS indicating the level of precision obtained in the calibration process. 

The stability of the instrument attitude and orbital control system (AOCS) is quantified in terms of mean performance error (MPE),
performance reproducibility error (PRE) and relative performance error (RPE). These are ESA defined pointing error terms\footnote{http://peet.estec.esa.int/files/ESSB-HB-E-003-Issue1(19July2011).pdf} affecting the measured timelines via mainly two mechanisms: 1) the drifting of the spectrum along the spectral
axis of the detector array, from here on referred to as ‘spectral jitter’; 2) the drift of the spectrum along
the spatial direction (or ‘spatial jitter’). The effect of jitter on the observed timelines is the
introduction of correlated noise, characterized by the power-spectrum of the telescope pointing.
The amplitude of the resultant photometric scatter depends on the amount of spectral/spatial
displacement of the spectrum, the PSF of the instruments, the detector intra-pixel  response and
the amplitude of the inter-pixel variations.
The effects of spectral jitter are not simulated by \echosim. This is motivated by the reasoning that drifts in the spectrum along the spectral axis of the array can be effectively removed during data reduction using the several stellar emission and absorption lines detected with high significance (see e.g. \cite{waldmann2012}\cite{waldmann2013}). 
The spatial component of the jitter affects the point source signal only and is simulated through a second-order variation between the position of the sampled PSF in the spatial direction in Equation~\ref{eq:pointsource} and the location of the detector pixel:
\begin{eqnarray*}
 Q_J(i,j,t) = && Q_P(i,j,t) \frac{1}{p_s} \left( f_{\rm eff}\frac{\partial p_s}{\partial y_j} <\delta \theta>_{\Delta T}\right. + \\
	  &&\left.+\frac{f_{\rm eff}^2}{2}\frac{\partial^2 p_s}{\partial y_j^2} <\delta \theta^2 >_{\Delta T}\right) ,
\end{eqnarray*}
where $<\delta \theta>_{\Delta T}$   is the time-averaged angular displacement of the telescope's line-of-sight from the target, and $\Delta T$ is the detector sampling interval. The jitter $\delta \theta$ is simulated as a Gaussian random process with an user-defined power spectrum (and bandwidth) which defines the electromechanical response of the AOCS.  
When the AOCS loop is closed on the information provided by a star sensor (in the case of \echo\ this is the FGS in Figure~\ref{fig:1}),  pointing information is also available. \echosim\ provides this housekeeping information with a user-defined accuracy on the absolute pointing knowledge and update rate. This information can be used by a data reduction pipeline to de-correlate the effects of pointing jitter on the reconstructed exoplanet spectrum. The reason to model the pointing jitter as a first and second order effect on the detected timelines is for computational efficiency. It would be more computationally  onerous to simulate the jitter early on in the process because of the different bandwidths characterizing the astronomical signal, the detector sampling rates and the AOCS. 

The combined output of the simulation 
\begin{eqnarray*}
 &&Q_{\rm tot}(i,j,t) =  \Delta T Q_{DC}(i,j,t) + Q_{RO}(i, j, t) + \\
     && \;\;G_{ij} \Delta T \left[ Q_P(i,j,t) + Q_D(i,j) + Q_O(i,j) + Q_J(i,j,t) \right]    
\end{eqnarray*}
is stored by the Output Module for spectral-reconstruction analyses.
\begin{figure}[t]
  \centering
    \includegraphics[width=0.4\textwidth]{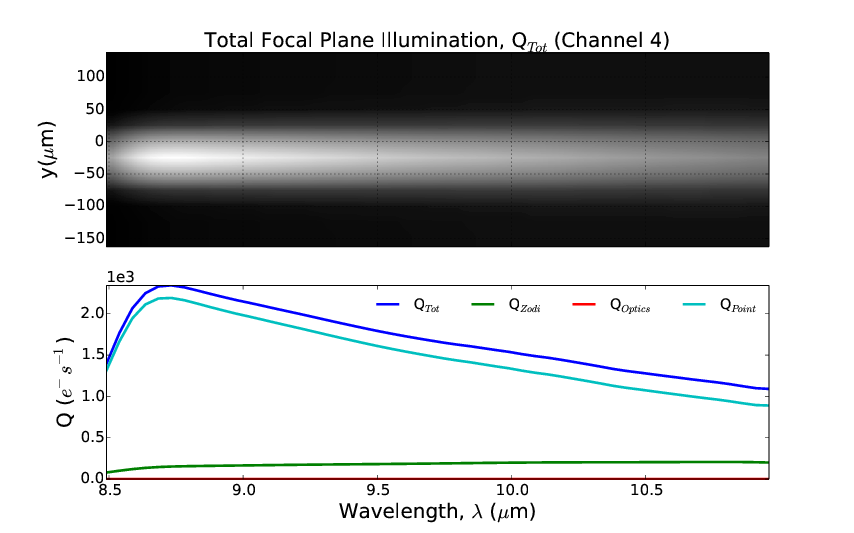} 
 
\caption{Snapshot of \echosim\ output for the MWIR-2 channel. Top: Focal plane illumination
including the science spectrum, instrumental and astrophysical backgrounds. Bottom: Plot of the
total and individually contributing signals. }
\label{fig:3}       
\end{figure}

\subsection{Output Module}
In the Output module each detector timeline, $Q_{\rm tot}$, is saved into FITS files. Each fits file is a frame,
i.e. data collected during one integration time, $\Delta T$. Each FITS file contains a number of images
extensions, each corresponding to one \echo\ channel. Additional housekeeping information is
also stored in the fits header. At the moment this includes only the FGS, but can include
spacecraft telemetry housekeeping when required.

\section{Use of \echosim}
\echosim\ simulates all aspects of the spectroscopic detection of the exoplanet atmosphere. The simulation is done in the time-domain, and includes all major systematic aspects known to challenge the confidence with which a detection is obtained. This is an invaluable tool in a wide range of experimental activities and scientific analyses. 

A data reduction pipeline is the first tool required in order to reconstruct the exoplanet's atmospheric spectrum from the raw simulated data. \echosim\ allows the development of novel data reduction pipelines which can be tested against the simulation's inputs. This allows the validation of the reduction pipeline  in reconstructing the signal in the presence of random correlated and uncorrelated noise sources and systematics. A data reduction pipeline has been developed to analyse \echosim\ simulations in the context of the \echo\ mission study. The pipeline is distributed along with \echosim\ and it is fully discussed in \cite{waldmann2014}. Case studies discussed in the next section have been analysed using this reduction software. 

\echosim\ is used to validate the ability of a proposed instrument concept and mission scenario to deliver its scientific goals. Similarly, the proposed scientific goals and mission scenario can be used by \echosim\ to obtain technical instrument requirements such as pointing stability, detector noise and stability, temperature stability, etc., required for a successful detection. Therefore this simulator plays a central role when designing an instrument or when formulating a mission concept as, for instance, \echo.

\echosim\ simulations of exoplanet primary and secondary eclipses are used by \cite{barstow2014} to demonstrate how  \echo\ would be effective in constraining atmospheric models using the methodology discussed in \cite{barstow2013a} and \cite{barstow2013b}. 

Spectroscopy of transiting planets has been pioneered in recent years with a number of ground-based and space instruments. The detection involves measuring the small modulation imposed by the atmosphere of the transiting planet over the much larger component of the parent star. Therefore, detections always involve controlling instrumental and astrophysical systematic effects at a level of one part in $10^3$--$10^4$ compared to the signal from the star. From this challenge, controversies about detections are likely to emerge. One possible way to demonstrate that systematics are under robust control at the required level is to use \echosim\ to simulate the detection. Thanks to its high modularity and instrument parametrization, \echosim\ can simulate virtually any instrument used in recent years to pioneer the field of transiting spectroscopy, and their observational strategies. Models of NICMOS, WFC3, on the {\it Hubble Space Telescope}, IRS and IRAC, on {\it Spitzer}, can be implemented and 
analyses repeated to validate claims of molecular detections on planets like HD\,189733\,b, HD\,209458\,b, GJ\,1214\,b, etc.

\section{Case Studies}
\begin{figure}[t]
  \centering
  \begin{minipage}[t][5.5cm][t]{0.4\textwidth}
    \includegraphics[width=\textwidth]{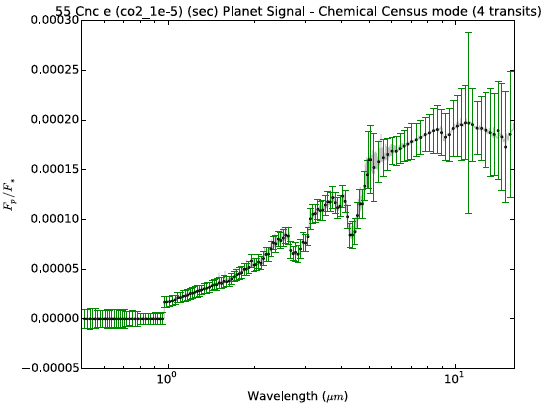}  
  \end{minipage} 
  \begin{minipage}[b]{0.4\textwidth}
    \includegraphics[width=\textwidth]{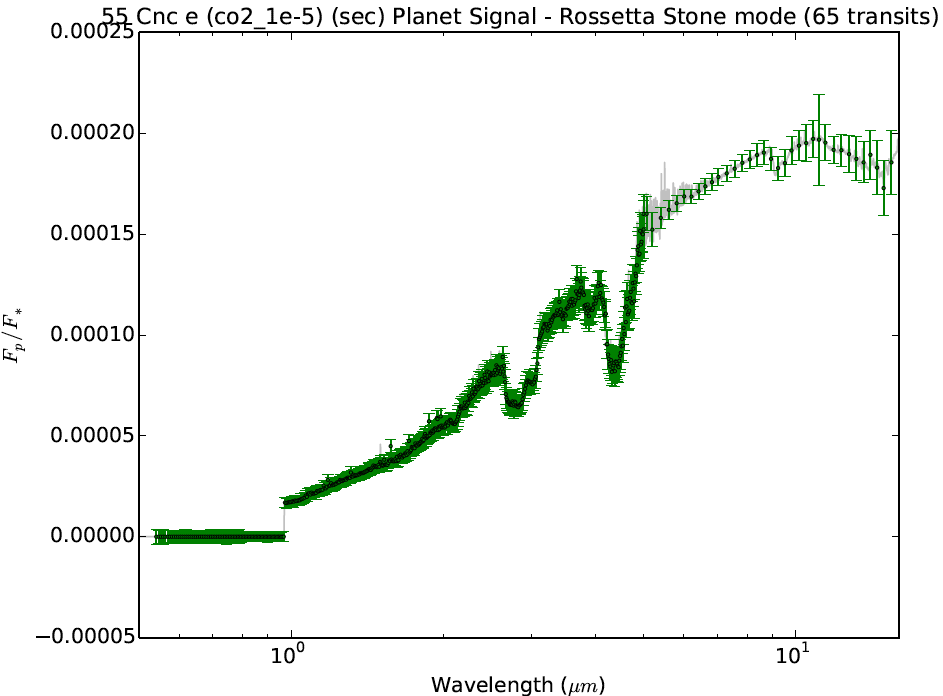}
   
  \end{minipage}
\caption{The spectra of the exoplanet 55 Cnc e are reconstructed to high significance. The top and bottom panels are representative of what can be achieved by co-adding several transits. These represent \echo's ``Census'' (top panel) and Rosetta-stone (bottom panel) observing modes.}
\label{fig:4}       
\end{figure}

\echosim\ was originally developed to assess the \echo\ mission concept using realistic time-domain simulations, which include all major critical aspects which are likely to give rise to the most challenging systematic effects. Figure~\ref{fig:4} shows an analysis done for the transiting hot super Earth 55\,Cnc\,e, at wavelengths longer than 1\micron. The simulations for secondary transits capture the fidelity with which the emission spectrum can be detected with a dedicated space mission, after combining either 4 or 65 transits. These correspond to the ``Chemical Census'' mode and ``Rosetta-stone modes'', respectively, discussed in the yellow book. Similar simulations  can be used to predict molecular detectability (see e.g. \cite{tessenyi2013}), to estimate observing times which, in turn, affect scheduling\cite{morales2014}, but also to show how \echo\ can detect different types of exoplanets\cite{varley2014}.

\begin{figure}[t]
  \centering
    \includegraphics[width=0.4\textwidth]{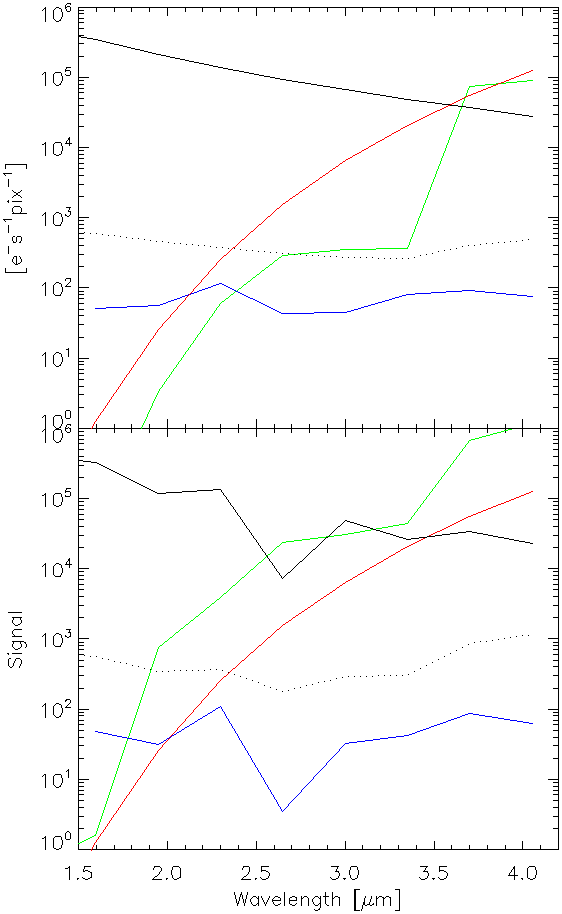}
   
\caption{\echosim\ can be used to study the capabilities of ground based (bottom panel) and balloon-borne (top panel) instruments in detecting a secondary eclipse of HD~189733~b. A low resolution, $R\sim 5$ spectrometer is used in these simulations. Black solid line: signal from the combined star and planet; green solid line: Earth atmospheric emission; blue solid line: exoplanet emission; red solid line: emission from optical surfaces; black dotted line: expected noise in one second of integration. }
\label{fig:5}       
\end{figure}

Stratospheric ballooning provides an effective way to conduct observations in a low atmospheric pressure environment. A telescope operating in the upper stratosphere, at an altitude of 38\,km, or higher, experiences observational conditions which are vastly improved compared to ground instrumentation. With flight durations from 2 weeks (Long Duration Ballooning)  to ~100 days (Ultra Long Duration Ballooning) operated by NASA's Columbia Scientific Balloon Facility, balloon platforms offer an effective alternative when similar observations from the ground are difficult or impossible. In Figure~\ref{fig:5} \echosim\ simulations investigate a possible balloon mission pivoting around a low-resolution, $R\sim 5$, spectrometer concept, operating in the NIR from 1.5 to 4\micron. No space instrument is currently operating over this wavelength region, which is actively investigated by ground telescopes. The two panels in the figure show identical information for a balloon mission (top panel) and for a ground-based 
instrument (bottom panel). Earth atmospheric 
emission (green line) at balloon altitude is reduced by several orders of magnitude compared to the ground case. The flux from the star+planet system and from the planet emission alone are shown by the black and blue solid lines, respectively. \echosim's noise estimated in one second of integration is also shown (black dotted line). This analysis shows how ground instrumentation struggles in the attempt to extract the exoplanet signal from the low signal-to-noise detection. Sky variability adds to the problem, but the vastly improved conditions at balloon altitude, combined with the expected stability of the stratosphere, would allow robust detection of  methane (e.g. \cite{swain2010}\cite{mandell2011}\cite{waldmann2012}), C$_2$H$_2$, HCN, etc. (e.g. \cite{tinetti2013}).

\section{Future Developments}
The \echosim\ project will continue to be maintained in the next few years, and two main additions are expected. A detailed detector model featuring detector non-idealities is planned to be implemented. This allows simulating {\it HST}-WFC3 and NICMOS and {\it Spitzer}-IRAC detections. It is known that the effects of stellar stability can be effectively decorrelated from the timelines when monitoring of the stellar flux in the optical is available. Also, the effect of stellar flux variations are less severe when observing at long IR wavelengths. However, this is an effect which can be easily implemented by \echosim\ and it would allow the investigation of how effectively faint targets can be detected, in particular warm to temperate Earth-size planets orbiting late type stars. 

\section{Conclusions}
We have implemented an end-to-end time-domain instrument simulator to study the most critical and challenging aspects involved in the detection and characterization of the atmospheres of extrasolar planets with the method of transit spectroscopy. The simulator capabilities extend beyond those a static radiometric model can provide by implementing time-varying instrumental effects as pointing jitter and time domain correlated and uncorrelated noise realisation. This is of particular importance because the method of the detection involves measuring small time modulations the transiting planet imposes on the much larger signal of the parent star, and a non-optimal control of the systematics result in a coupling of the star signal with the exoplanet signal, completely dazzling it. This is particularly severe because of the interplay between intra-pixel and inter-pixel response and pointing jitter. In this paper, we have presented a detailed description of the algorithms implemented by \echosim\ to deliver realistic 
simulations. The software is very efficient, with full simulations run in seconds on modern laptop computers, hence enabling the possibility of Monte Carlo analyses. \echosim\ has been used to assess the \echo\ space mission, and can be adopted as a tool to develop novel mission concepts, or to investigate detection confidence for existing sub-orbital and space instrumentation.

%

%


\bibliography{echosim}   

\bibliographystyle{spphys}       

\end{document}